\documentclass[11pt]{article}
\usepackage{hyperref}
\pdfoutput=1
 \usepackage{hyperref}
 \pdfoutput=1
\begin{document}
\title{ Current-Driven Flow across a Stationary Jellyfish}
\author{Christina Hamlet$^1$, Laura A. Miller$^2$, \\ Roger Fan$^2$, Makani Dollinger$^2$, Steven Harenber$^2$\\
\\\vspace{6pt} $^1$ Department of Mathematics, North Carolina State University,\\ $^2$ Department of Mathematics, University of North Carolina at Chapel Hill}
\maketitle
\begin{abstract}
We present several dye visualization and numerical simulation fluid dynamics videos of a sessile jellyfish subjected to channel flow. The \href{file:////anc/HamletV038LowRes.mp4}{low resolution video} and the \href{file:////anc/HamletV038HighRes.mp4}{high resolution video} display the vortex patterns in different channel flows. This description accompanies the video submission V038 to the 2011 APS DFD Gallery of Fluid Motion.
\end{abstract}
\section{Introduction}
We simulate Cassiopea (upside-down) jellyfish in channel flow using a classical immersed boundary method as developed by Charles Peskin and using DataTank (Visual DataTools, Inc.) for post processing of images. We performed laboratory dye visualization using a flow tank constructed by Dr. Steve Vogel (Duke University Dept. of Biology) and fluorescein dye (Flinn Scientific. Lot number: 118372.)
 A full pulse of the simulated jellyfish consists of a cycle of a contraction (0.6 s), a short pause (0.13 s), a relaxation (0.7 s) and a longer pause (2.0 s.) Oscillating and periodic one-directional flows have a frequency of 0.21 Hz. Channel flow speeds were simulated at 0.0 cm/s, 1.0 cm/s and 2.0 cm/s as indicated in the videos. \\
\indent Each numerical simulation shows a vorticity plot for the fluid around the bell of the model jellyfish. The simulations show the speed and direction of the current influence the formation and shedding of vortices around the tips of the bell. The simulations also indicate the prominent oral arms structures play a role in vorticity trapping and shedding. Dye visualization experiments qualitatively reveal similar results.
\end{document}